\documentclass{elsart5}

 \usepackage{graphicx}

\usepackage{amssymb}
\newcommand{\mib}[1]{\mbox{\boldmath $#1$}}

\begin{document}

\begin{frontmatter}

\title{Theoretical Analysis of the Reduction of N\'{e}el Temperature in La$_{2}$(Cu$_{1-x}$Zn(or Mg)$_x)$O$_{4}$}

\author{T. Edagawa\corauthref{cor1}}
\ead{oguchi@rs.noda.tus.ac.jp}
\corauth[cor1]{}
\author{Y. Fukumoto and A. Oguchi}

\address{Department of Physics, Faculty of Science and Technology, 
Tokyo University of Science, 2641 Yamazaki, Noda, Chiba 278-8510, Japan}

\begin{abstract}
Using Tyablikov's decoupling approximation, we calculate the initial suppression rate of the N\'{e}el temperature,
$R_{\rm{IS}}=-\lim_{x\rightarrow 0} T^{-1}_{\rm N}$ $dT_{\rm N}/dx$,
in a quasi two-dimensional diluted Heisenberg antiferromagnet with nonmagnetic impurities of concentration $x$.
In order to explain an experimental fact that $R^{(\rm {Zn})}_{\rm{IS}}=3.4$ of the Zn-substitution is different from
$R^{(\rm {Mg})}_{\rm{IS}}=3.0$ of the Mg-substitution, we propose a model in which impurity substitution
reduces the intra-plane exchange couplings surrounding impurities, as well as dilution of spin systems.
The decrease of $12$\% in exchange coupling constants by Zn substitution and decrease of $6$\% by Mg substitution
explain those two experimental results, when an appropriate value of the interplane coupling is used.
\end{abstract}


\end{frontmatter}


The discovery of high-$T_{\rm c}$ compounds renewed interests in quasi two-dimensional quantum antiferromagnets.
The site dilution problem of two-dimensional quantum antiferromagnets has been extensively studied.~\cite{Cheong,Kato,Chernyshev,Mucciolo,Castro}
In this paper, we treat a problem of the difference between Zn and Mg substitutions in La$_{2}$CuO$_{4}$.~\cite{Cheong}

As far as the present authors know, on the impurity problem of square-lattice Heisenberg antiferromagnets
nonmagnetic impurities have been treated just as static vacancies in spin systems in all previous theoretical studies .
However, in an experiment on La$_{2}$(Cu$_{1-x}$Zn(or Mg)$_x)$O$_{4}$, Cheong {\it et al.} reported that
the initial suppression rate of N\'{e}el temperature
\begin{equation}
   R_{\rm{IS}}=-\lim_{x\rightarrow 0} \frac{1}{T_{\rm N}} \frac{dT_{\rm N}}{dx}
\end{equation}
shows impurity ion dependence, i.e., 
$R^{\rm{(Zn)}}_{\rm{IS}}=3.4$ for Zn substitution and $R^{\rm{(Mg)}}_{\rm{IS}}=3.0$ for Mg substitution.~\cite{Cheong}
This result indicates that the nonmagnetic impurities are not a role of static vacancies in the spin system.
Cheong {\it et al.} also measured the orthorhombic-to-tetragonal transition temperature $T_{O\leftrightarrow T}$ of 
La$_{2}$(Cu$_{1-x}$Zn(or Mg)$_x)$O$_{4}$, and found a faster increase of $T_{O\leftrightarrow T}$ of Zn substitution than Mg substitution.
On the basis of this observation, they concluded that Mg is a better substantial ion than Zn for realizing spin dilution systems.~\cite{Cheong}

We write the Hamiltonian of the parent compound without impurities as
\begin{equation}
   H = J \sum_{\langle i,j \rangle_{\parallel}} \mib{S}_{i}\cdot\mib{S}_{j}
         +J_{\perp} \sum_{\langle i,j \rangle_{\perp}} \mib{S}_{i}\cdot\mib{S}_{j},
\end{equation}
where $i$ and $j$ denote lattice points on a simple cubic lattice, 
$\langle i,j \rangle_{\parallel}$ are nearest-neighbor pairs of sites on a square-lattice plane, 
and $\langle i,j \rangle_{\perp}$ are inter-plane nearest neighbor pairs.
Although inter-plane exchange couplings in La$_{2}$CuO$_{4}$ are more complex than those in the above model,
we simply treat the three dimensionality within the form of the above Hamiltonian.
An ``ideal" nonmagnetic impurity is introduced at site $O$, then the spin system is perturbed by
\begin{equation}
   H'_{O} = -J \sum_{\langle O, j \rangle_{\parallel}} \mib{S}_{O}\cdot\mib{S}_{j}
         -J_{\perp} \sum_{\langle O,j \rangle_{\perp}} \mib{S}_{O}\cdot\mib{S}_{j}.
\end{equation}
The Hamiltonian of the diluted spin system is given by $H_{\rm{tot}}=H+\sum_{O}H'_{O}$, where the sum runs over
$N x$ lattice points chosen at random from the total $N$ lattice points.
On the basis of the Tyablikov decoupling approximation, MacGurn calculated configuration averaged
Green's functions with self-energy corrections within first order in $x$, and estimated $R_{\rm{IS}}$
for a  square lattice case, $J_{\perp}\rightarrow 0$, and for an isotropic simple cubic lattice case, $J_{\perp}=J$.~\cite{MacGurn}
MacGurn's results are $R_{\rm{IS}}(J_{\perp}\rightarrow 0)=\pi$ and $R_{\rm{IS}}(J_{\perp}=J)=1.68$.
The former value for the square lattice is close to the experimental values for La$_{2}$(Cu$_{1-x}$Zn(Mg)$_x)$O$_{4}$,
but other mechanisms are needed to get a quantitative agreement between theory and experiment.
One of such mechanisms is a weak inter-plane coupling.

\begin{figure}[t]
\begin{center}
\includegraphics[width=.75\linewidth]{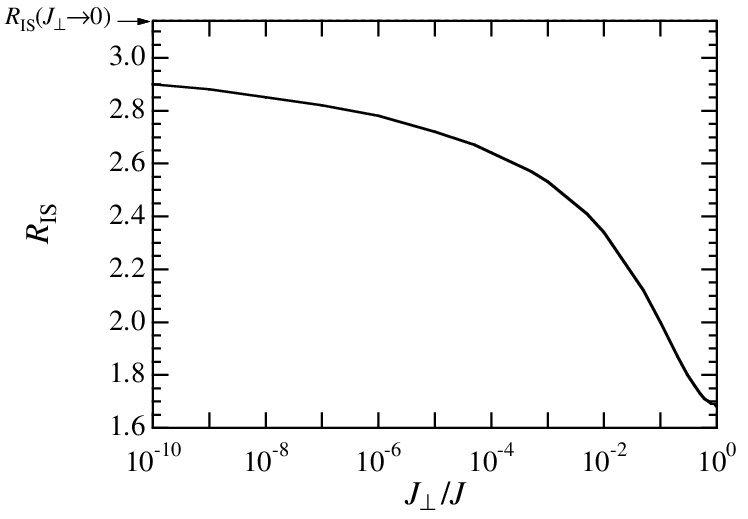}
\end{center}
\caption{Initial suppression rate $R_{\rm IS}$ as a function of 
interplane coupling $J_{\perp}$.}
\label{fig-1}
\end{figure}

We begin with extending MacGurn's calculation to study how $R_{\rm{IS}}$ depends on $J_{\perp}$.
The result is shown in Fig.~\ref{fig-1}. We find that $R_{\rm{IS}}$ decreases monotonically as 
$J_{\perp}$ increases. This behavior is reasonable because the three dimensionality is expected to
stabilize antiferromagnetic ordering. If we assume $J_{\perp}=5.0 \times 10^{-5} J$ in La$_{2}$CuO$_{4}$,~\cite{Hucker} 
then we obtain $R_{\rm{IS}}=2.67$.
This value is much smaller than $R^{\rm{(Zn)}}_{\rm{IS}}=3.4$ and $R^{\rm{(Mg)}}_{\rm{IS}}=3.0$,
and thus we have to search for an additional mechanism which destabilizes the antiferromagnetic ordering and makes $R_{\rm{IS}}$ increase.

\begin{figure}[b]
\begin{center}
\includegraphics[width=1\linewidth]{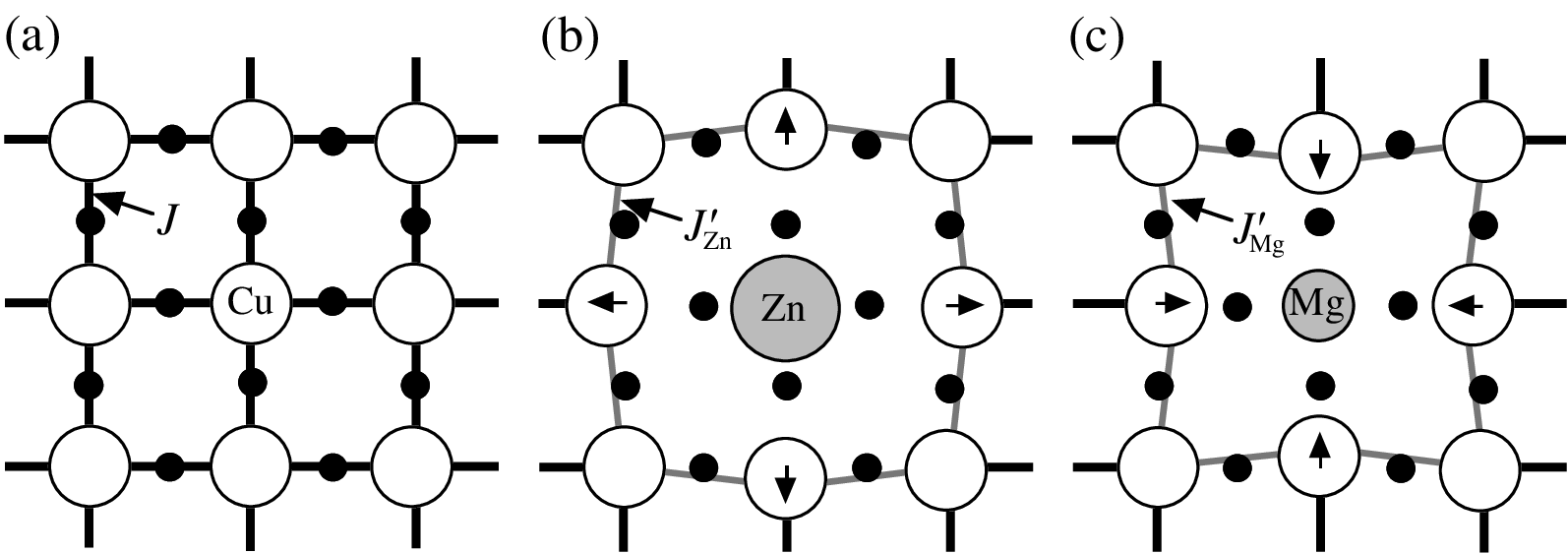}
\end{center}
\caption{Schematic representations of (a) a CuO$_2$ plane, where
the small closed circles represent positions of oxygens, (b) a CuO$_2$ plane 
with Zn-substitution, and (c) a CuO$_2$ plane with Mg-substitution.}
\label{fig-2}
\end{figure}

The experimental result of $R^{\rm{(Zn)}}_{\rm{IS}}\neq R^{\rm{(Mg)}}_{\rm{IS}}$ gives us a hint in searching for the additional mechanism. 
A natural way to understand this experimental result is that the exchange couplings around nonmagnetic impurities
are changed depending on the ion radius of the impurities. 
The ion radii of Zn, Cu, Mg are, respectively, 0.74, 0.73, 0.72 ð.~\cite{Cheong}
When a Cu ion substitutes for a Zn ion, then the four nearest neighbor Cu ions of the impurity
moves away from the impurity and in the case of Mg they approach the impurity, as shown in Fig.~\ref{fig-2}.
(We neglect smaller displacements of other Cu ions.)
Then, for eight exchange bonds surrounding the impurity ion, the linear arrangement of lobes of two $d_{x^2-y^2}$ orbitals 
and a $p_{\sigma}$ orbital is changed to a distorted one,
which leads to the reduction of overlap integral between such two Cu sites, i.e., $t \rightarrow t-\Delta t$.
The reduction of the overlap integral also gives $J \rightarrow J'=J(1-2\Delta t/t)$ for small $\Delta t/t$.

\begin{figure}[t]
\begin{center}
\includegraphics[width=.75\linewidth]{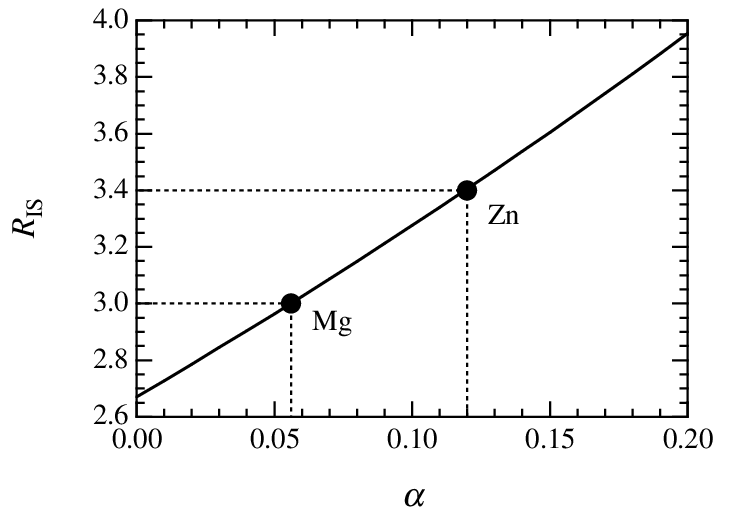}
\end{center}
\caption{Initial suppression rate $R_{\rm IS}$ as a function of $\alpha$
for $J_{\perp}/J=5.0 \times 10^{-5}$.}
\label{fig-3}
\end{figure}

In order to incorporate this effect, we write the perturbation due to an impurity at site $O$ as 
\begin{equation}
   \tilde{H}'_{O} = H'_{O}-J \alpha \sum_{\langle i, j \rangle_O} \mib{S}_{i}\cdot\mib{S}_{j},
\end{equation}
where $\alpha=2\Delta t/t$, and the sum runs over eight intra-plane nearest-neighbor pairs surrounding the impurity site $O$.
Following MacGurn's procedure, we calculate $R_{\rm{IS}}$ for the Hamiltonian $\tilde{H}_{\rm{tot}}=H+\sum_{O}\tilde{H}'_{O}$.
In the limit $J_{\perp}\rightarrow 0$, we obtain $R_{\rm{IS}} \simeq \pi+A \alpha$ with a rather large coefficient $A=32-16\pi+5\pi^2/2\simeq 6.4$.
The calculated result for $J_{\perp} = 5.0 \times 10^{-5} J$, which corresponds to La$_{2}$CuO$_{4}$, is shown in Fig.~\ref{fig-3}.
Using this result and experimental results of $R^{({\rm{Mg}})}_{\rm{IS}}=3.0$ and $R^{({\rm{Zn}})}_{\rm{IS}}=3.4$,
we obtain $\alpha^{({\rm{Zn}})} = 0.12$ and $\alpha^{({\rm{Mg}})} = 0.06$ for La$_{2}$(Cu$_{1-x}$Zn(Mg)$_x)$O$_{4}$,
which means that reductions of the overlap integral due to Zn and Mg substitutions are, respectively, 6\% and 3\%.
The result of $\alpha^{({\rm{Zn}})}>\alpha^{({\rm{Mg}})}$ shows that Mg is a better substantial ion than Zn,
which is consistent with the experimental finding by Cheong {\it et al.} mentioned before.~\cite{Cheong}

In summary, we conclude that the reduction of intra-plane exchange couplings around impurities
causes a rather fast increase of $R_{\rm{IS}}$ and gives the origin of the difference in $R_{\rm{IS}}$
between La$_{2}$(Cu$_{1-x}$Zn$_x)$O$_{4}$ and La$_{2}$(Cu$_{1-x}$Mg$_x)$O$_{4}$.
As a future problem, 
it is also interesting to incorporate four-spin ring-exchange terms into our model,~\cite{Coldea}
which is expected to get the estimations of $\alpha$ to decrease more or less because of frustration effects.~\cite{Lauchli}


\begin{thebibliography}{00}
\bibitem{Cheong} S-W.~Cheong {\it et al.}: Phys. Rev. B{\bf 44} (1991), p.~9739.
\bibitem{Kato} K.~Kato {\it et al.}: Phys. Rev. Lett. {\bf 84} (2000), p.~4204.
\bibitem{Chernyshev} A.~L.~Chernyshev {\it et al.}: Phys. Rev. B{\bf 65} (2002), p.~104407.
\bibitem{Mucciolo} E.~R.~Mucciolo {\it et al.}: Phys. Rev. B{\bf 69} (2004), p.~214424.
\bibitem{Castro} E.~V.~Castro {\it et al.}: Phys. Rev. B{\bf 73} (2006), p.~054422.
\bibitem{MacGurn} R.~MacGurn: J. Phys. C{\bf 27} (1979), p.~3523.
\bibitem{Hucker} M.~Hucker {\it et al.}: Phys. Rev. B \textbf{59} (1999) p.~725.
\bibitem{Coldea} R.~Coldea {\it et al.}: Phys. Rev. Lett. \textbf{86} (2001) p.~5377.
\bibitem{Lauchli} A.~L\"{a}uchli {\it et al.}: Phys. Rev. Lett. \textbf{95} (2005) p.~137206.
\end{thebibliography}
\end{document}